\documentclass[12pt]{iopart}

\usepackage{iopams}
\usepackage{graphicx}
\expandafter\let\csname equation*\endcsname\relax
\expandafter\let\csname endequation*\endcsname\relax
\usepackage{amsmath}
\usepackage{bm}
\usepackage{hyperref}
\usepackage{graphics}
\usepackage{slashed}
\usepackage{amssymb}
\usepackage{url}
\usepackage[dvipsnames,svgnames,table]{xcolor}
\usepackage[T1]{fontenc}

\makeatletter
\newcommand{\tpitchfork}{%
  \vbox{
    \baselineskip\z@skip
    \lineskip-.52ex
    \lineskiplimit\maxdimen
    \m@th
    \ialign{##\crcr\hidewidth\smash{$-$}\hidewidth\crcr$\pitchfork$\crcr}
  }%
}
\makeatother

\usepackage[normalem]{ulem}
\usepackage{color}

\begin{document}

\title[Covariant Dynamical Systems Formulation of the TOV Equations]{Covariant Dynamical Systems Formulation of the Tolman–Oppenheimer–Volkoff Equations}
\author{Eduardo Bittencourt${}^{1,2}$\footnote{Corresponding author} and Mariam Campbell${}^{3}$ and Peter K. S. Dunsby${}^{2,4,5}$ and Sergio E. Jor\'as${}^{6}$}
\address{${}^{1}$Institute of Physics and Chemistry, Federal University of Itajub\'a, Av.\ BPS 1303, Itajub\'a/MG, Brazil}
\address{${}^{2}$Department of Mathematics and Applied Mathematics, University of Cape Town, Rondebosch 7700, Cape Town, South Africa}
\address{${}^{3}$Astrophysics Research Centre, School of Agriculture and Science, University of KwaZulu-Natal, Private Bag X54001, Durban 4000, South Africa}
\address{${}^{4}$Center for Space Research, North-West University, Potchefstroom 2520, South Africa}
\address{${}^{5}$South African Astronomical Observatory, Observatory 7925, Cape Town, South Africa}
\address{${}^{6}$Instituto de F\'\i sica, Universidade Federal do Rio de Janeiro,\\
 CEP 21941-972 Rio de Janeiro, RJ, Brazil}
\ead{bittencourt@unifei.edu.br, CampbellM@ukzn.ac.za, peter.dunsby@uct.ac.za, joras@if.ufrj.br}

\date{\today}

\begin{abstract}
We revisit static, spherically symmetric perfect-fluid stellar models in General Relativity within the framework of the $1+1+2$ semi-tetrad formalism. For locally rotationally symmetric static spacetimes, the Tolman-Oppenheimer-Volkoff system can be expressed as a covariant first-order dynamical system and, after suitable normalization, reformulated as a three-dimensional autonomous flow for a general equation of state (EoS). In the case of a linear EoS, the system reduces further to a planar dynamical system whose finite and asymptotic equilibrium points, together with their stability properties, admit a clear geometrical interpretation in terms of covariant variables. For more general equations of state, such as the polytropic case, the dynamics naturally acquire a genuinely three-dimensional character. Beyond providing a compact, covariant, and physically transparent reformulation of the relativistic stellar problem, the present analysis clarifies how the standard metric description is encoded within a global phase-space structure constructed from geometrically meaningful covariant variables.
\end{abstract}

\vspace{2pc}

\maketitle

\section{Introduction}

Compact stars remain one of the primary laboratories for testing relativistic gravity in the strong-field regime. In the standard metric formulation, hydrostatic equilibrium is governed by the Tolman-Oppenheimer-Volkoff (TOV) equations \cite{Tolman1939,OppenheimerVolkoff1939}. Beyond direct numerical integration, these equations have also been investigated using qualitative and dynamical-systems techniques, which reveal the global structure of the stellar solution space in terms of invariant sets, separatrices, and asymptotic states \cite{ChavezNamboSarbach2021}. Early work along these lines was carried out by Collins \cite{Collins1985}, and later developed systematically for relativistic stars with linear and polytropic equations of state by Nilsson and Uggla, and by Heinzle, R\"ohr, and Uggla \cite{NilssonUggla2000,NilssonUggla2000Polytropic,HeinzleRohrUggla2003}.

A complementary perspective is provided by the $1+1+2$ covariant formalism, which is particularly well adapted to static, spherically symmetric spacetimes and reformulates the stellar problem as a first-order system for geometrically meaningful scalar variables \cite{Clarkson2007,CarloniVernieri2018,NaiduCarloniDunsby2021}. This framework arises from a further decomposition of the standard $1+3$ covariant approach with respect to a preferred spatial direction. For locally rotationally symmetric class II geometries, all vector and tensor degrees of freedom vanish, and the dynamics are completely encoded in a small set of covariant scalars \cite{vanElstEllis1996}. One of the principal advantages of the formalism is that it preserves covariance while remaining algebraically equivalent to the standard metric description. The $1+1+2$ approach therefore provides a natural bridge between the traditional metric formulation and the qualitative analysis of stellar solution spaces.

The aim of this letter is to formulate the General Relativity (GR) stellar problem within the $1+1+2$ framework from a qualitative and dynamical-systems perspective. On the one hand, this yields a compact reformulation of the TOV system in terms of covariantly defined variables with direct geometrical interpretation. On the other hand, it exposes the global organization of the solution space in a manner that renders invariant sets, critical structures, and distinguished stellar trajectories transparent.

We first examine the linear EoS, for which the reduced system becomes polynomial and admits a complete local and global analysis. We then contrast this with the polytropic case, where the loss of functional homogeneity prevents the same planar reduction and instead gives rise to a three-dimensional autonomous flow. In this way, the present formulation connects naturally with the existing dynamical-systems literature on relativistic stars and relativistic polytropic models \cite{Collins1985,NilssonUggla2000,NilssonUggla2000Polytropic,HeinzleRohrUggla2003,AnderssonBurtscher2019}.
Throughout this letter, units are chosen such that $c=8\pi G=1$.

\section{Covariant stellar system}\label{sec:cov_star}

For locally rotationally symmetric class II static spacetimes, the $1+1+2$ formalism is constructed from a timelike unit vector $u^a$ together with a preferred radial spacelike unit vector $e^a$ orthogonal to it \cite{Clarkson2007}. In the isotropic static case, the relevant covariant scalars are the radial acceleration $\mathcal A \equiv e^a \dot{u}_a$, the expansion of the two-dimensional sheets $\phi \equiv \delta_a e^a$, the electric Weyl scalar $\mathcal E$, and the matter variables $\rho$ and $p$. Here, an overdot denotes the covariant derivative along $u^a$, while $\delta_a$ represents the projected covariant derivative on the two-dimensional space orthogonal to both $u^a$ and $e^a$. Denoting derivatives along $e^a$ by $\hat X = e^a D_a X$, the stellar equilibrium equations take the form \cite{CarloniVernieri2018,NaiduCarloniDunsby2021}
\begin{align}
\hat{\phi} &= -\frac{1}{2}\phi^2+\mathcal A\phi-\rho-p,
\label{eq:phihat}\\
\hat p &=-(\rho+p)\mathcal A,
\label{eq:phat}\\
\hat{\mathcal A} &=-(\mathcal A+\phi)\mathcal A+\frac{1}{2}(\rho+3p),
\label{eq:Ahat}
\end{align}
supplemented by the constraint
\begin{equation}
\mathcal E=-\mathcal A\phi+\frac13(\rho+3p).
\label{eq:constraint}
\end{equation}
These equations are equivalent to the usual TOV system \cite{CarloniVernieri2018}.

Introducing the dimensionless radial variable
\[
\zeta \equiv 2\ln\frac{r}{r_0},
\]
it is convenient to rewrite the system in terms of derivatives with respect to $\zeta$ (denoted by a prime), using the relation $X'=\hat X/\phi$. We then introduce the normalized variables
\begin{equation}
\label{eq:norm_chart}
Y=\frac{\mathcal A}{\phi}, \qquad
\tilde K=\frac{K}{\phi^2}, \qquad
\tilde\mu=\frac{\rho}{\phi^2}, \qquad
\tilde p=\frac{p}{\phi^2}, \qquad
\Xi=\frac{\phi'}{\phi},
\end{equation}
with $K$ being the Gaussian curvature of the 2-sheet, introduced for later convenience. Rewriting the dynamical system \eqref{eq:phihat}-\eqref{eq:Ahat} in terms of this new set, one obtains
\begin{align}
\tilde K'&=-2\tilde K\left(\tilde K-\tilde\mu-\frac14\right),
\label{eq:Kred}\\
\tilde p'&=-\tilde p^2+\tilde p\left(\tilde\mu-3\tilde K+\frac74\right)
-\tilde\mu\left(\tilde K-\frac14\right).
\label{eq:pred}
\end{align}
Two constraints were used to eliminate $Y$ and $\Xi$, namely
\begin{equation}
Y=\tilde K+\tilde p-\frac14,
\qquad
\Xi=\tilde K-\tilde\mu-\frac34.
\label{eq:constraints-normalized}
\end{equation}
Thus, whenever the EoS yields a closure relation of the form $\tilde\mu=\tilde\mu(\tilde p)$, the stellar problem reduces to an autonomous planar system. 

\section{Linear EoS}

In the case of linear EoS, the physical and normalized variables give the exact same closure
\begin{equation}
\rho=\lambda p\quad\Longrightarrow\quad\tilde\mu=\lambda \tilde p,
\label{eq:linearclosure}
\end{equation}
where $\lambda\equiv {\rm const.}$ is the EoS parameter. 
Substituting Equation~\eqref{eq:linearclosure} into Equations \eqref{eq:Kred} and \eqref{eq:pred}, we obtain the closed system
\begin{align}
\tilde K'&=-2\tilde K\left(\tilde K-\lambda\tilde p-\frac14\right),
\label{eq:sys1}\\
\tilde p'&=\tilde p\left[(\lambda-1)\tilde p-(\lambda+3)\tilde K+\frac{\lambda+7}{4}\right].
\label{eq:sys2}
\end{align}
Although the phase-space analysis can be carried out on the full real line, the physically reasonable range corresponds to $\lambda\geq1$, otherwise the system may experience mechanical or thermodynamical instabilities.

When one of the components of the vector field defined from the dynamical system vanishes, we have the nullclines. In this case, when $\tilde K'=0$, we have
\begin{equation}
\tilde K=0,
\quad {\rm or}
\qquad
\tilde K=\lambda\tilde p+\frac14,
\label{eq:Knull}
\end{equation}
and when $\tilde p'=0$, we find
\begin{equation}
\tilde p=0,
\quad
\qquad {\rm or}
\qquad
\tilde K=\frac{\lambda-1}{\lambda+3}\tilde p+\frac{\lambda+7}{4(\lambda+3)},
\qquad (\lambda\neq -3).
\label{eq:pnull}
\end{equation}
Their intersections yield the finite equilibrium points, denoted by $P_{i}:(\tilde{K}_{i},~\tilde{p}_{i})$, as follows
\begin{align}
P_0&:(0,0), \nonumber\\
P_{1/4}&:\left(\frac14,0\right), \nonumber\\
P_{\rm int}&:\left(\frac14+\frac{\lambda}{(1+\lambda)^2},
\frac{1}{(1+\lambda)^2}\right), \qquad (\lambda\neq -1), \nonumber\\
P_{K0}&:\left(0,-\frac{\lambda+7}{4(\lambda-1)}\right), \qquad (\lambda\neq 1).
\label{eq:fixedpoints}
\end{align}

After linearization, finding the stability of the fixed points is straightforward: the point $P_0$ is a source for $\lambda>-7$ and a saddle for $\lambda<-7$, while $P_{1/4}$ is always a saddle. For $P_{K0}$, the eigenvalues of the Jacobian are
\begin{equation}
l_1=-\frac{\lambda^2+6\lambda+1}{2(\lambda-1)},
\qquad
l_2=-\frac{\lambda+7}{4},
\label{eq:eigsPK0}
\end{equation}
whereas for $P_{\rm int}$ it is better to evaluate in terms of the trace $\tau_\star$ and determinant $\Delta_\star$ at $P_{\rm int}$, as follows
\begin{equation}
\tau_\star=-\frac{\lambda+3}{2(\lambda+1)},
\qquad
\Delta_\star=\frac{\lambda^2+6\lambda+1}{2(\lambda+1)^2}.
\label{eq:traceint}
\end{equation}
Defining $\lambda_\pm=-3\pm 2\sqrt2$, one finds that $P_{K0}$ is a source for $\lambda<-7$, a sink whether $\lambda_-<\lambda<\lambda_+$ or $\lambda>1$, and a saddle otherwise, while $P_{\rm int}$ is a saddle for $\lambda_-<\lambda<\lambda_+$ and a sink otherwise.

A structural property of Equations \eqref{eq:sys1}--\eqref{eq:sys2} is that both $\tilde K=0$ and $\tilde p=0$ are invariant sets. Therefore, the 2D flow cannot cross quadrants. In particular, the sector $\tilde K,\tilde p>0$ is the most natural configuration to associate with compact stars.

Furthermore, since the closure is linear, the system is quadratic at large radii in phase space. More generally, if
\begin{equation}
\lambda_\infty:=\lim_{|\tilde p|\to\infty}\frac{\tilde\mu(\tilde p)}{\tilde p},
\end{equation}
then the leading homogeneous part of the flow is
\begin{align}
\tilde K'&\sim -2\tilde K^2+2\lambda_\infty \tilde K\tilde p,
\label{eq:asympK}\\
\tilde p'&\sim (\lambda_\infty-1)\tilde p^2-(\lambda_\infty+3)\tilde K\tilde p.
\label{eq:asympp}
\end{align}
For the linear EoS, we can identify $\lambda_\infty=\lambda$. Thus, the Poincar\'e compactification places the plane onto the unit disk via 
\[
(\tilde K,\tilde p)\mapsto (U,V)=\frac{(\tilde K,\tilde p)}{\sqrt{1+\tilde K^2+\tilde p^2}},
\]
where the boundary $U^2+V^2=1$ represents directions at infinity (for more details, see Ref.\ \cite{DumortierLlibreArtes2006}, for instance). For the quadratic leading system \eqref{eq:asympK}--\eqref{eq:asympp}, the equilibria on the circle can be determined using polar coordinates $\tilde{K}=r\cos\theta$ and $\tilde p=r\sin\theta$ and searching for fixed points of the angular drift $\theta$ at infinity
\begin{equation}
    \theta'=\frac{\tilde K\tilde p'-\tilde p \tilde K'}{r^2}\sim -\frac{(\lambda_{\infty}+1)\tilde K \tilde p (\tilde K + \tilde p)}{r^2}.
\end{equation}
Therefore, universal directions are given by
\begin{equation}
\theta'=0\Longrightarrow 
\begin{cases}
\tilde p=0:\theta=0,\pi\Rightarrow (U,V)=(\pm1,0),\\[1ex]
\tilde K=0:\theta=\frac{\pi}{2},\frac{3\pi}{2}\Rightarrow (U,V)=(0,\pm1),\\[1ex]
\tilde p=-\tilde K:\theta=\frac{3\pi}{4},\frac{7\pi}{4}\Rightarrow (U,V)=(\pm\frac{1}{\sqrt{2}},\mp\frac{1}{\sqrt{2}}),
\end{cases}
\label{eq:infty_directions}
\end{equation}
i.e. three antipodal pairs on the boundary for the generic case $\lambda_\infty\neq \pm 1$.

The linearization of the desingularized compactified flow is straightforward using the stereographic projections, yielding the eigenvalue pairs and their corresponding stability (up to orientation conventions) summarized in Table \ref{tab:fixed_points_alpha}. It should also be remarked that some equilibria at infinity become non-hyperbolic when $\lambda_\infty=\pm 1$ and require higher-order terms, which is beyond the scope of this paper.

\begin{table}[ht]
\centering
\begin{tabular}{|c|c|c|c|}
\hline
\textbf{Asymptotic Fixed point} & $\boldsymbol{\lambda_\infty<-1}$ & $\boldsymbol{-1<\lambda_\infty<1}$ & $\boldsymbol{\lambda_\infty>1}$ \\
\hline
$E_{\tilde p=0}:\left(2,\,-\lambda_\infty-1\right)$ 
& Source/Sink & Saddle & Saddle \\
\hline
$E_{\tilde K=0}:\left(1-\lambda_\infty,\,\lambda_\infty+1\right)$ 
& Saddle & Source/Sink & Saddle \\
\hline
$E_{\tilde p=-\tilde K}:\left(2\lambda_\infty+2,\,\lambda_\infty+1\right)$ 
& Sink/Source & Source/Sink & Source/Sink\\
\hline
\end{tabular}
\caption{Stability of the fixed points as a function of $\lambda_\infty$.}
\label{tab:fixed_points_alpha}
\end{table}
  
\begin{figure}[htb]
    \centering
    \includegraphics[width=10cm,height=10cm]{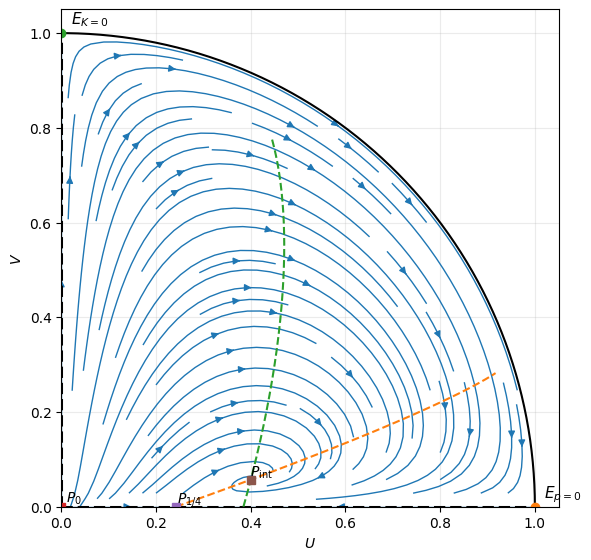}
    \caption{Radiation case ($\lambda=3$), in compactified coordinates $\{U,V\}=(1+\tilde K^2+\tilde p^2)^{-\frac12}\{\tilde K,\tilde p\}$. At the physically relevant sector ($\tilde K,\tilde p\geq0$), the intersections of nullclines (dashed lines) give the equilibria $P_0$ (source), $P_{1/4}$ (saddle) and $P_{\rm int}$ (stable focus), including the axes. At infinity, the fixed points $E_{K=0}$ and $E_{p=0}$ are both saddle.}
    \label{fig:linear}
\end{figure}

As an illustration, Figure \ref{fig:linear} shows the phase portrait for the radiation case $\lambda=3$ in the physically relevant first quadrant. The finite equilibria are $P_0$ (source), $P_{1/4}$ (saddle), and $P_{\rm int}$ (stable focus), while the points at infinity $E_{\tilde p=0}$ and $E_{\tilde K=0}$ are both saddles. In this sector there is a unique distinguished orbit, namely the unstable separatrix of $P_{1/4}$ connecting it to $P_{\rm int}$. This is precisely the physical trajectory associated, up to self-similar transformations, with the Misner-Zapolsky solution \cite{MisnerZapolsky1964} discussed by Collins \cite{Collins1985}. All other trajectories in this quadrant terminate at curvature singularities at finite radius \cite{Collins1985}. It should be mentioned that the phase portraits for any $\lambda\geq1$ will be qualitatively the same as the one shown in Figure \ref{fig:linear}.

\section{Polytropic equation of state}

We now consider the polytropic EoS
\begin{equation}
p=\kappa \rho^\gamma, 
\label{eq:polyEOS}
\end{equation}
where $\gamma=1+1/n$, with $n$ being the polytropic index, and $\kappa>0$. Unlike the linear EoS, the polytropic relation is not homogeneous of degree one in $(\rho,p)$. As a consequence, the qualitative analysis depends more strongly on the choice of variables. In the normalized chart (\ref{eq:norm_chart}), the stellar problem is naturally recast as a three-dimensional autonomous flow, which is well suited to expose invariant sets, nullclines, and the global phase-space structure. In the original covariant variables (\ref{eq:phihat})-(\ref{eq:Ahat}), by contrast, the same system gives a more transparent physical description of the stellar interior, especially the role of the center, the monotonicity of the matter variables, and the identification of the surface. The two formulations are therefore best viewed as complementary.

\subsection{Normalized formulation}

By rewriting Equation (\ref{eq:polyEOS}) in terms of the normalized variables, one finds
\begin{equation}
\tilde p=\kappa \phi^{2(\gamma-1)}\tilde\mu^\gamma,
\label{eq:normalizedpoly}
\end{equation}
which depends explicitly on $\phi$. Hence, the planar reduction used in the linear case is no longer available: there is no autonomous closure of the form $\tilde\mu=\tilde\mu(\tilde p)$ or $\tilde p=\tilde p(\tilde\mu)$ unless $\gamma=1$.

To keep the connection with the linear system as transparent as possible, it is convenient to introduce the dimensionless ratio 
\begin{equation}
w\equiv\frac{p}{\rho}=\kappa \rho^{1/n},
\label{eq:sigmaDef}
\end{equation}
so that $\tilde\mu=\tilde p/w$ and $c_s^2=\gamma w$. 

In this way, the normalized GR system may be written directly in terms of
$(\tilde K,\tilde p,w)$ as
\begin{align}
\tilde K'&=-2\tilde K\left(\tilde K-\frac{\tilde p}{w}-\frac14\right), \label{eq:polySystem1}\\
\tilde p'&=\tilde p\left[\left(\frac{1}{w}-1\right)\tilde p-\left(\frac{1}{w}+3\right)\tilde K+\frac{1+7w}{4w}\right], \label{eq:polySystem2}\\
w'&=-\frac{1+w}{n+1}\left(\tilde K+\tilde p-\frac14\right). \label{eq:polySystem3}
\end{align}
Thus, the polytropic problem is naturally described by a three-dimensional autonomous flow in $(\tilde K,\tilde p,w)$. The advantage of this form is that the first two equations are structurally the same as in the linear case after the formal replacement $\lambda\longrightarrow 1/w$, the difference being that $1/w$ is now itself a dynamical variable rather than a constant EoS parameter. 

Since a more mathematical treatment of relativistic polytropes has already been given in Ref.~\cite{NilssonUggla2000Polytropic}, we restrict the attention here to the physically relevant region
\[
\tilde K\geq 0,\qquad \tilde p\geq 0,\qquad w\geq 0.
\]

The nullclines are now surfaces in $(\tilde K,\tilde p,w)$-space. From Equations \eqref{eq:polySystem1}-\eqref{eq:polySystem2}, one obtains that, for each fixed $w$, the $\tilde K$- and $\tilde p$-nullclines have exactly the same algebraic form as in the linear case, with $\lambda\longrightarrow1/w$. Finally, from Equation~\eqref{eq:polySystem3}, we get
\begin{equation}
\mathcal N_{w}^{(1)}:\ w=-1,
\qquad
\mathcal N_{w}^{(2)}:\ \tilde K+\tilde p=\frac14.
\label{eq:sigmanullpoly}
\end{equation}
In the physical sector $w>0$, only the second of these is relevant.

As the fixed points are intersections of the nullclines, we see that there are no finite interior fixed points in the physical sector. Indeed, the only equilibrium touching the physical sector is the line
\begin{equation}
\mathcal L:\qquad
(\tilde K,\tilde p,w)=\left(\frac14,0,w_\star\right),
\qquad
w_\star\ \text{arbitrary},
\label{eq:vacuumline}
\end{equation}
which reflects the fact that $w=\tilde p/\tilde\mu$ is not intrinsically defined when
$\tilde p,\tilde\mu\rightarrow 0$. Note that the values of $\tilde K$ and
$\tilde p$ at $\mathcal L$ coincide with those of $P_{1/4}$ in the linear case. Thus, as in the previous formulation, this line should be interpreted
as a degenerate representation of the regular stellar center. Its linear stability is
\begin{equation}
\operatorname{spec}(J_{\mathcal L})
=
\left\{-\frac12,\ 1,\ 0\right\},
\end{equation}
so $\mathcal L$ is a hyperbolic saddle line. 

\begin{figure}[htb]
    \centering
    \includegraphics[width=\textwidth,height=15cm]{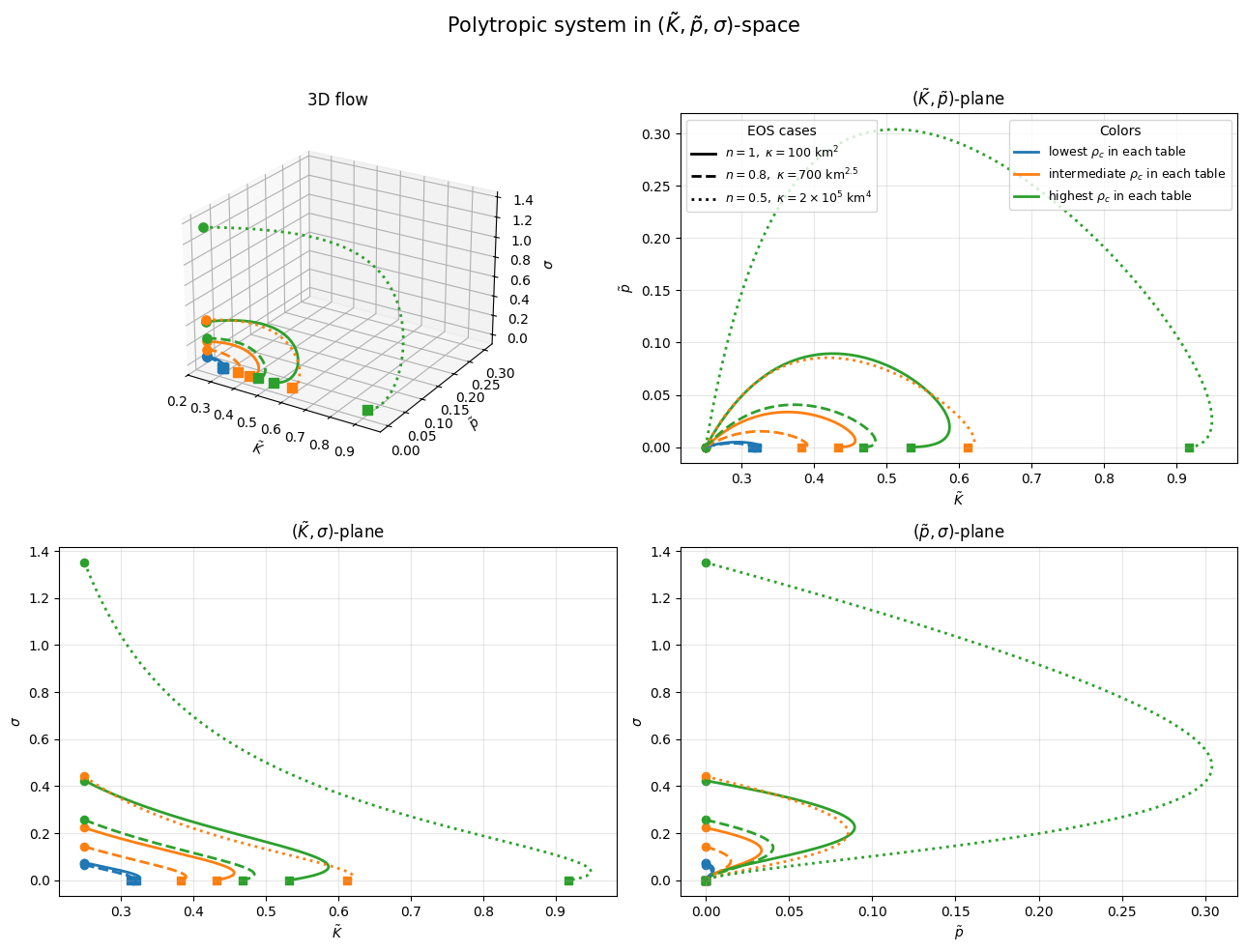}
    \caption{Polytropic case in the normalized variables $(\tilde K,\tilde p,w)$, for representative stellar models taken from Kokkotas and Ruoff \cite{KokkotasRuoff2001}. The three families correspond to $n=1$, $\kappa=100\,\mathrm{km}^2$ (solid lines), $n=0.8$, $\kappa=700\,\mathrm{km}^{2.5}$ (dashed lines), and $n=0.5$, $\kappa=2\times10^5\,\mathrm{km}^{4}$ (dotted lines). Within each family, the colors indicate increasing central density: for $n=1$, $\rho_c=\{1.0,\,3.0,\,5.7\}\times10^{15}\,\mathrm{g\,cm^{-3}}$; for $n=0.8$, $\rho_c=\{0.8,\,1.5,\,2.4\}\times10^{15}\,\mathrm{g\,cm^{-3}}$; and for $n=0.5$, $\rho_c=\{0.8,\,2.0,\,3.5\}\times10^{15}\,\mathrm{g\,cm^{-3}}$. Circles mark the initial point near the degenerate center line, while squares indicate the stellar surface, where $w\to0$.}
    \label{fig:polytropic}
\end{figure}

By using the constraint \eqref{eq:constraints-normalized}, Equation~\eqref{eq:polySystem3} can be written as
\begin{equation}
w'=-\frac{1+w}{n+1}\,Y.
\label{eq:sigmaprimeY}
\end{equation}
Hence the sign of $w'$ is controlled entirely by the sign of $Y$. Moreover, from the normalized covariant equation for $Y$, one finds on the hypersurface $Y=0$,
\begin{equation}
Y'\big|_{Y=0} = \frac12\left(\tilde\mu+3\tilde p\right),
\end{equation}
which is positive for $\tilde p>0$ and $w>0$. Therefore, the domain
\begin{equation}
\mathcal D_+:= \left\{ \tilde K>0,\ \tilde p>0,\ w>0,\ Y>0 \right\}
\end{equation}
is forward invariant. Furthermore, on $\mathcal D_+$, the function
\begin{equation}
V(w)=\ln(1+w)
\end{equation}
is strictly monotonic, since
\begin{equation}
V'=-\frac{Y}{n+1}<0.
\label{eq:monotoneV}
\end{equation}
Therefore, $V$ defines a strict monotonic function on the physically relevant invariant region, thereby excluding the existence of periodic or recurrent orbits within that sector of the state space. Nevertheless, $V$ does not constitute a Lyapunov function for any equilibrium point of the original three-dimensional system, since the equilibrium line $\mathcal L$ possesses saddle character.
The principal strength of the normalized formulation is geometric in nature. It renders the global organization of the polytropic flow explicit through its nullclines, invariant boundaries, and monotonic structure, while at the same time preserving a close analogy with the linear system via the effective parameter $1/w$. Its main limitation, however, is that the regular stellar center is no longer represented by a distinguished finite interior point of the state space.
Figure~\ref{fig:polytropic} illustrates the physically relevant sector of the normalized polytropic flow in the variables $(\tilde K,\tilde p,w)$ for representative models adapted from Kokkotas and Ruoff \cite{KokkotasRuoff2001}. The trajectories originate near the degenerate center line, where $\tilde K\to 1/4$, $\tilde p\to 0$, and $w\to w_c$, and evolve toward the low-pressure boundary defined by $w\to 0$. In the $(\tilde K,\tilde p)$ projection, the trajectories closely resemble the phase portrait of the linear-EoS system, reflecting the fact that Equations~\eqref{eq:polySystem1}--\eqref{eq:polySystem2} retain the same underlying algebraic structure as in the linear case. The additional variable $w$ then quantifies the departure from this effectively linear behavior and governs the drift of the trajectories through the full three-dimensional state space. Increasing the central density drives the orbits farther from the weak-field regime, while smaller values of the polytropic index $n$ produce larger excursions in all three dynamical variables.

\subsection{Original covariant formulation}

The same polytropic problem can also be analyzed directly in the original $1+1+2$ system \eqref{eq:phihat}-\eqref{eq:Ahat}, regarding $\rho$ as a function of $p$. The stellar equations then form a closed autonomous system in the variables $(\phi,\mathcal{A}, p)$, with the hat derivative playing the role of the evolution parameter along the preferred radial direction.

The nullclines are now surfaces in $(\phi,\mathcal{A},p)$-space. From Equation~\eqref{eq:phihat}, we get
\begin{equation}
\hat\phi=0
\quad\Longleftrightarrow\quad
\mathcal{A}=\frac{\phi}{2}+\frac{\rho(p)+p}{\phi},
\label{eq:null_phi_hat_explicit}
\end{equation}
for $\phi\neq 0$. From Equation~\eqref{eq:phat}, we obtain
\begin{equation}
\hat p=0
\quad\Longleftrightarrow\quad
\mathcal{A}=0,
\label{eq:null_p_hat}
\end{equation}
in the physical sector. Finally, from Equation~\eqref{eq:Ahat}, we have
\begin{equation}
\hat{\mathcal A}=0
\quad\Longleftrightarrow\quad
\frac{-\phi\pm\sqrt{\phi^2+2(\rho(p)+3p)}}{2}.
\label{eq:null_A_hat_explicit}
\end{equation}
Since $\rho+3p\geq 0$, the square root is always real in the physical region.

Note that the vacuum subset $\mathcal{V}:=\{p=0,\ \rho=0\}$ is invariant. Indeed, on $p=0$ one has $\rho=0$ for the polytropic EoS, and it gives $\hat p=0$. Restricted to $\mathcal{V}$, the system becomes planar:
\begin{align}
\hat\phi &= -\frac12\phi^2+A\phi,
\label{eq:vacuum_phi_hat}\\
\hat{\mathcal A} &= -(\mathcal{A}+\phi)\mathcal{A}.
\label{eq:vacuum_A_hat}
\end{align}

Regarding the finite fixed points, they must simultaneously satisfy $\hat\phi=\hat p=\hat{\mathcal A}=0$. In the physical region, the only fixed point is the origin $(\phi,\mathcal{A},p)=(0,0,0)$, which is not the regular stellar center. A regular center corresponds instead to
\begin{equation}
r\to 0:\qquad
\phi\sim \frac{2}{r}\to\infty,\qquad
\mathcal{A}\to 0,\qquad
p\to p_c,\qquad
\rho\to \rho_c,
\label{eq:regular_center_hat}
\end{equation}
and is therefore located at infinity in the $(\phi, \mathcal{A}, p)$-chart. Thus, the physical stellar solutions are not organized by finite interior equilibria, but by a one-parameter family of regular center
data $(p_c,\rho_c)$.

The main advantage of the original variables is that the monotonic behavior of the stellar profiles becomes immediate. At a regular center, we have $\mathcal{A}=0$, and from Equation~\eqref{eq:Ahat}, we get
\begin{equation}
\hat{\mathcal A}\big|_{\mathcal{A}=0}=\frac12(\rho+3p)>0
\qquad
(p>0).
\label{eq:A_grows_from_center}
\end{equation}
Hence, $\mathcal{A}$ becomes positive immediately outside the center. Moreover, it cannot cross back through zero while the matter variables remain positive, because at any point with $\mathcal{A}=0$ and $p>0$ one again has $\hat{\mathcal A}>0$. 

Regarding the pressure, by using Equation~\eqref{eq:phat}, it follows that
\begin{equation}
\hat p=-(\rho+p)\mathcal{A}<0
\qquad\text{for}\qquad
p>0.
\label{eq:p_monotone_hat}
\end{equation}
So, the pressure decreases strictly along the outward radial direction. The same holds for the density, since $\hat\rho=\hat p/c_s^2<0$. Thus, every regular stellar model is represented by a monotonic orbit starting from the regular center data at $\phi=\infty$, $\mathcal{A}=0$, $p=p_c$, and evolving toward the first point where
\begin{equation}
p(R)=0.
\label{eq:surface_hat}
\end{equation}
This identifies the stellar surface as the first intersection of the orbit with the invariant vacuum
subset $\mathcal{V}$.

The strength of the original covariant formulation is therefore mainly physical. It makes the center-to-surface flow transparent: the center is encoded by the asymptotic regime $\phi\to\infty$,
the gravitational acceleration $\mathcal{A}$ is positive throughout the interior, and the pressure and density decrease monotonically until the orbit reaches the vacuum boundary. The price paid for this transparency is that the regular center is no longer represented by a finite point of the autonomous system.

Taken together, the two formulations highlight complementary aspects of the same stellar problem. The normalized variables are better suited to the global phase-space geometry, whereas the original covariant variables are better suited to the direct interpretation of the stellar interior.

\section{Contact with the metric approach and previous results}
\label{sec:metric-literature}

To connect the covariant formulation with the standard description of relativistic stellar structure, consider the static, spherically symmetric line element in Schwarzschild-like coordinates,
\begin{equation}
ds^2=-e^{2\Phi(r)}dt^2+\left(1-\frac{2m(r)}{r}\right)^{-1}dr^2+r^2d\Omega^2.
\label{eq:metric-ansatz}
\end{equation}
For spherical symmetry, the Gaussian curvature of the 2-sheets is $K=1/r^2$, and the $1+1+2$ variables are related to the metric functions by
\begin{equation}
\phi=\frac{2}{r}\sqrt{1-\frac{2m}{r}},
\qquad
\mathcal{A}=\Phi_{,r}\sqrt{1-\frac{2m}{r}},
\qquad
K=\frac{1}{r^2},
\label{eq:covariant-metric-map}
\end{equation}
with the derivatives related through
\begin{equation}
\hat{X}=\sqrt{1-\frac{2m}{r}}\,\frac{dX}{dr}.
\label{eq:hat-derivative-metric}
\end{equation}
The normalized variables introduced in Section~\ref{sec:cov_star} then take the form
\begin{equation}
Y=\frac{r\Phi_{,r}}{2},
\qquad
\tilde K=\frac{1}{4\left(1-\frac{2m}{r}\right)},
\qquad
\tilde\mu=\frac{r^2\rho}{4\left(1-\frac{2m}{r}\right)},
\qquad
\tilde p=\frac{r^2p}{4\left(1-\frac{2m}{r}\right)}.
\label{eq:normalized-metric-map}
\end{equation}
Thus, the covariant variables retain a direct physical interpretation: $Y$ measures the logarithmic gradient of the redshift potential, $\tilde K$ is a normalized compactness variable, and $\tilde\mu$ and $\tilde p$ are the density and pressure rescaled by the sheet expansion. It should be noted that fixed points of the normalized variables are achieved at finite values of the parameter $r$, the ``evolution'' parameter of the dynamics.

This map also makes it straightforward to recover the usual TOV system. Substituting Equation~\eqref{eq:normalized-metric-map} into the constraint $Y=\tilde K+\tilde p-\frac14$, one obtains
\begin{equation}
\Phi_{,r} = \frac{2m+r^3p}{2r(r-2m)}.
\label{eq:metric-potential-gradient}
\end{equation}
Likewise, Equation~\eqref{eq:Kred} yields
\begin{equation}
\frac{dm}{dr}=\frac12\,\rho r^2,
\label{eq:mass-equation-ourunits}
\end{equation}
while Equation~\eqref{eq:phat} becomes
\begin{equation}
\frac{dp}{dr}=-(\rho+p)\Phi_{,r}.
\label{eq:hydrostatic-step}
\end{equation}
Combining Equations \eqref{eq:metric-potential-gradient} and \eqref{eq:hydrostatic-step}, we recover the TOV equation in the units adopted here,
\begin{equation}
\frac{dp}{dr}=-\frac{(\rho+p)(2m+r^3p)}{2r(r-2m)}.
\label{eq:TOV-ourunits}
\end{equation}
Therefore, the $1+1+2$ system is an exact covariant rewriting of the standard metric formulation, as already shown in Ref.\ \cite{CarloniVernieri2018}.

The same metric map also translates classical compactness inequalities directly into the covariant variables. In particular, since Equation (\ref{eq:normalized-metric-map}) expresses $\tilde K$ algebraically in terms of $m/r$, Buchdahl’s bound \cite{Buchdahl1959} for a regular isotropic fluid sphere with nonincreasing density immediately gives
\begin{equation}
\frac{2M}{R}\le \frac89,
\end{equation}
where $M$ is the total mass $M$ and $R$ is the surface radius, which implies the compactness inequality
\begin{equation}
\tilde K(R)\le \frac94.
\end{equation}
Thus, the normalized curvature at the stellar surface cannot grow without bound. Since $Y=\tilde K+\tilde p-\frac14$ and $\tilde p(R)=0$ for an ordinary fluid surface, one also has $Y(R)\leq 2$.

More generally, Buchdahl's refined inequality
\begin{equation}
1-\frac{2M}{R}\ge \frac19(1+\beta)^2,    
\end{equation}
gives
\begin{equation}
\tilde K(R)\le \frac{9}{4(1+\beta)^2},    
\end{equation}
where $\beta$ is an auxiliary parameter encoding an \emph{a priori} bound on the local pressure-to-density ratio, defined by $\rho/(3p)\ge \beta^{-1}$. In particular, $\beta=1$ corresponds to the condition $T^\mu{}_\mu\ge 0$. 

This correspondence also clarifies the relation between the present analysis and earlier dynamical-systems studies. For the linear equation of state $\rho=\lambda p$, the exact closure relation $\tilde\mu=\lambda\tilde p$ reduces the stellar equations to the planar polynomial system \eqref{eq:sys1}--\eqref{eq:sys2}. The resulting invariant sets, critical points, and asymptotic directions reproduce the qualitative structures identified in previous analyses of relativistic stellar models. In particular, when $\lambda=3$, corresponding to the radiation case, the phase space contains a unique physical orbit in the first quadrant, namely the unstable separatrix of $P_{1/4}$, which is associated with the Misner-Zapolsky solution discussed by Collins \cite{Collins1985}. Similarly, the emergence of a higher-dimensional autonomous system in the polytropic case is consistent with the enlarged state spaces introduced by Nilsson and Uggla, and by Heinzle, R\"ohr, and Uggla \cite{NilssonUggla2000,NilssonUggla2000Polytropic,HeinzleRohrUggla2003}.
A principal advantage of the present formulation is that it combines qualitative transparency with direct geometrical interpretation. In the standard TOV approach, the global organization of the solution space is encoded in a non-autonomous radial system and is therefore comparatively less transparent. By contrast, the normalized covariant variables employed here render invariant sectors, nullclines, critical structures, and asymptotic behavior explicit, while remaining algebraically linked to the metric functions and matter variables. In this sense, the $1+1+2$ framework provides a natural bridge between the traditional metric description and the qualitative analysis of relativistic stellar models.

\section{Conclusions and outlook}

We have presented a compact GR benchmark for static, spherically symmetric perfect-fluid stars within the $1+1+2$ covariant formalism. After normalization, the covariant TOV equations reduce to an autonomous planar system whenever a barotropic closure relation is available. For the linear EoS, this leads to an exact polynomial flow whose finite critical points, invariant sets, and asymptotic directions can be determined analytically in full. The radiation case illustrates the power of the formalism especially clearly: the global phase portrait not only isolates the unique physical separatrix, but also reveals the dynamical origin and global role of the corresponding stellar solution. For polytropes, the qualitative structure changes fundamentally. The loss of homogeneity prevents the planar closure of the system, and the natural dynamical description becomes intrinsically three-dimensional. In this setting, the physically admissible region contains no finite interior equilibrium points, yet it admits a strict monotonic function that strongly constrains the global flow. The contrast between the linear and polytropic cases therefore clarifies which qualitative properties arise directly from the GR closure relations and which instead require a genuinely higher-dimensional dynamical framework.

One of the central outcomes of the present analysis is that the $1+1+2$ variables are not simply an alternative parametrization of the stellar problem, but a covariantly defined set of geometrical quantities that remain directly connected to the standard metric formulation. Through the map defined by Equation~(\ref{eq:covariant-metric-map}), the autonomous system can be translated algebraically into the usual mass, pressure, density, and redshift functions. As a result, the qualitative structures identified in the phase space admit an immediate physical interpretation in terms of the conventional stellar variables. The formalism therefore provides not only a compact covariant reformulation of the TOV equations, but also a transparent bridge between the traditional metric description and the global phase-space perspective on relativistic stellar models. In this sense, the present GR analysis furnishes a natural benchmark and conceptual foundation for future qualitative studies of more general relativistic stellar systems.

\section*{Acknowledgements}
EB is partially supported by \textit{Conselho Nacional de Desenvolvimento Científico e Tecnológico} (grant N.\ 305217/2022-4) and FAPEMIG (APQ-05207-23). PKSD is supported by a grant from First Rand Bank (SA). SEJ is partially supported by FAPERJ.

\section*{References}

\bibliography{ref.bib}
\end{document}